\documentclass[12pt]{article}
\usepackage{latexsym}
\usepackage{epsfig,amssymb,euscript}
\usepackage{amsmath}
\numberwithin{equation}{section}  %DARIO: added this command
\newsavebox{\ns}
\newsavebox{\dbrane}
\def\be{\begin{equation}}
\def\ee{\end{equation}}
\def\bea{\begin{eqnarray}}
\def\eea{\end{eqnarray}}

\newcommand{\nn}{\nonumber}

\def\Dslash{\,\,{\raise.15ex\hbox{/}\mkern-12mu D}}
\def\Dbarslash{\,\,{\raise.15ex\hbox{/}\mkern-12mu {\bar D}}}
\def\delslash{\,\,{\raise.15ex\hbox{/}\mkern-9mu \partial}}
\def\delbarslash{\,\,{\raise.15ex\hbox{/}\mkern-9mu {\bar\partial}}}
\def\pslash{\,\,{\raise.15ex\hbox{/}\mkern-9mu p}}
\def\calDslash{\,\,{\raise.15ex\hbox{/}\mkern-12mu {\cal D}}}

\newcommand\R{\mathbb{R}}
\newcommand\Z{\mathbb{Z}}
\newcommand\F{\mathbb{F}}

\newcommand\C{\mathbb{C}}
\newcommand\T{\mathbb{T}}
\newcommand\diff{\mathrm{d}}
\newcommand\ex{\mathrm{e}}
\newcommand{\de}{\partial}
\newcommand{\vol}{\mathrm{vol}}

\begin{document}
\begin{titlepage}
\begin{center}
\today

\vskip 1.8cm
{\Large \bf  Notes on toric  Sasaki-Einstein \\[3mm]
seven-manifolds and AdS$_4$/CFT$_3$}

\vskip 1.5cm
{Dario Martelli$^{1*}$ ~~and~~ James Sparks$^{2}$}\\
\vskip 1.2cm

1: {\em Institute for Advanced Study\\
Einstein Drive, Princeton, NJ 08540, U.S.A.}\\
\vskip 0.8cm
2: {\em Mathematical Institute, University of Oxford,\\
 24-29 St Giles', Oxford OX1 3LB, U.K.}\\

\vskip 2.2cm

\end{center}

\vskip 1.5cm

\begin{abstract}
\noindent
We study the geometry and topology of two infinite families $Y^{p,k}$ of
Sasaki-Einstein seven-manifolds, that are expected to be  AdS$_4$/CFT$_3$
dual to families of ${\cal N}=2$ superconformal field theories in three dimensions.
These manifolds, labelled by two positive integers $p$ and $k$,
are Lens space bundles $S^3/\Z_p$ over $\C P^2$ and $\C P^1\times \C P^1$, respectively.
The corresponding Calabi-Yau cones are toric. We present their toric diagrams and gauged linear
sigma model charges in terms of $p$ and $k$, and find that the  $Y^{p,k}$ manifolds
interpolate between certain orbifolds of the homogeneous spaces $S^7, M^{3,2}$ and $Q^{1,1,1}$.
\end{abstract}

\vfill
\hrule width 5cm
\vskip 5mm

{\noindent $^*$ {\small On leave from: \emph{Blackett Laboratory,
Imperial College, London SW7 2AZ, U.K.}}}

\end{titlepage}
\pagestyle{plain}
\setcounter{page}{1}
\newcounter{bean}
\baselineskip18pt

%%%%%%%%%%%%%%%%%%%%%%%%%%%%%%%%%%%%%%%%%%%%%%%%%%%%%

\tableofcontents

\section{Introduction}

There has recently been renewed interest in supersymmetric three-dimensional
conformal field theories in the context of the AdS/CFT correspondence \cite{maldacena}.
The reasons for this interest are diverse. One motivation is that three-dimensional
CFTs describe the low-energy world-volume theory of coincident M2-branes. Until recently,
the understanding of these theories had been rather rudimentary. 
However, a breakthrough was made in the  work of \cite{BL}, where an
${\cal N}=8$ supersymmetric Chern-Simons theory was constructed.  
The authors of \cite{BL} proposed that this theory is related to the theory on coincident M2-branes. 
A careful study of the vacuum moduli space \cite{VMS} subsequently led to a more precise 
interpretation of this theory as describing two M2-branes at an $\R^8/\Z_2$ orbifold singularity.
A number of papers have further studied the proposal of \cite{BL}, culminating in the results
of \cite{ABJM} (ABJM). 
In the latter reference, the theory of \cite{BL} was recast in terms of an 
$SU(2)\times SU(2)$ Chern-Simons quiver gauge theory, which allowed for a generalisation\footnote{
More precisely, in this generalisation the gauge group is taken to be $U(N)\times U(N)$.} of the 
construction to an arbitrary number $N$ of M2-branes, with Chern-Simons level $k$.
The authors of \cite{ABJM} also discussed the gravity duals of these theories, 
showing that they are  AdS$_4\times S^7/\Z_k$  backgrounds of M-theory, with $N$ units of four-form flux.
These works open the way for a systematic study of AdS$_4$ M-theory backgrounds in terms of 
three-dimensional conformal field theories, using the AdS/CFT correspondence.

The analogous problems in the context of type IIB string theory are understood rather well. In this case the
gauge theories arise as the low-energy limit of D3-brane world-volume theories.
The maximally supersymmetric case is the ${\cal N}=4$ SYM theory. One can obtain
${\cal N}=1$ SCFTs by placing the D3-branes at a Calabi-Yau singularity. In the case of orbifolds or toric singularities
the technology to construct these gauge theories is now standard. The gravity duals of these theories are
type IIB AdS$_5\times Y_5$ backgrounds,  where $Y_5$ is a Sasaki-Einstein five-manifold \cite{Kehagias,KW,acharya,MP} (or
orbifold) with $N$ units of five-form flux.
For some time an obstacle  in the study of AdS$_5$/CFT$_4$ duals
was the lack of examples -- specifically, there existed
only two (non-orbifold) examples where the metric was known explicitly,  namely $S^5$ and $T^{1,1}$.
The discovery of the $Y^{p,q}$ Sasaki-Einstein 5-manifolds
in \cite{paper1,paper2} radically improved this situation.

In dimension seven, the classification of manifolds $Y_7$, not locally isometric to 
$S^7$, admitting Killing spinors 
falls into 3 types: weak $G_2$ manifolds, Sasaki-Einstein manifolds, and tri-Sasakian manifolds. 
These admit 1, 2 and 3 Killing spinors, respectively.
The Killing spinor equation immediately implies that 
the metric is Einstein with positive Ricci curvature, and that AdS$_4\times Y_7$, 
with $N$ units of four-form flux, is a supersymmetric solution to eleven-dimensional 
supergravity. The AdS/CFT dual theories then have $\mathcal{N}=1,2$ and $3$ supersymmetry, respectively. 
The metric cones $\diff r^2 + r^2 \diff s^2_7$ are Ricci-flat and are correspondingly $Spin(7)$, Calabi-Yau 
and hyper-K\"ahler cones, respectively. 
We note that toric tri-Sasakian manifolds are extremely well-studied -- see, for example, 
\cite{BG}. In particular, the Einstein metric on a toric tri-Sasakian manifold 
is the induced metric one obtains from a hyper-K\"ahler quotient construction.

In this paper we focus on Sasaki-Einstein manifolds that are not tri-Sasakian. 
In dimension seven,  the list of known explicit Sasaki-Einstein manifolds
in the literature, before \cite{paper3}, consisted of
the following: $M^{3,2}, Q^{1,1,1}$ and $V_{5,2}$.
For a review of these manifolds see, for example, \cite{DNP}.
The manifolds $M^{3,2}$ and $Q^{1,1,1}$ are natural generalisations
of $T^{1,1}$ in dimension five.  In particular, the corresponding Calabi-Yau cones are toric.
Proposals for the AdS/CFT duals of these homogeneous Sasaki-Einstein seven-manifolds
were given in \cite{tatar,italians,moreitalians}.
In \cite{paper3} the construction of \cite{paper1,paper2} was generalised
to arbitrary dimension, thus providing  infinite families
of Sasaki-Einstein manifolds in all odd dimensions.
The main result of \cite{paper3} shows that for any positive
curvature K\"ahler-Einstein manifold $B_{2n}$ there is a countably
infinite class of associated Sasaki-Einstein manifolds $Y_{2n+3}(B_{2n})$
\footnote{This construction has been subsequently
generalised in \cite{morese} to the case
where $B_{2n}$ is a product of K\"ahler-Einstein manifolds.}.
Here we will analyse two families in seven dimensions,
where $B_{4}$ is either $\C P^2$ or $\C P^1 \times \C P^1$. The seven-dimensional Sasaki-Einstein manifolds
will be denoted $Y^{p,k}(\C P^2)$ and $Y^{p,k}(\C P^1\times \C P^1)$, respectively.
Following  \cite{toric}, we will give a presentation
of the Calabi-Yau cones in terms of a  K\"ahler quotient, also known as a gauged linear sigma
model description \cite{wittenLSM}. In fact, note that the results of this analysis were 
anticipated in \cite{toric} (see the introduction of the latter reference).

The  description of the toric Calabi-Yau cones associated to the $Y^{p,q}$ metrics, presented in  \cite{toric},
gave important clues that aided the identification of the dual gauge theories. In particular, the
Calabi-Yau singularity is  (part of) the moduli space of supersymmetric vacua of these gauge theories.
It was also observed
that the family of $Y^{p,q}$ singularities interpolates between two limiting cases: $\C^3/\Z_{2p}$ and a $\Z_p$
orbifold of the conifold. For these the gauge theories are simple orbifolds of the ${\cal N}=4$ SYM and Klebanov-Witten
theories, respectively.
The geometric information in \cite{toric} was then used in  \cite{quivers} (see also \cite{BBC}) to identify
the general  family of quiver gauge theories.
Moduli spaces of orbifolds of the ABJM theory are currently under investigation 
 \cite{Benna:2008zy,Imamura:2008nn,Terashima:2008ba,Aharony:2008gk}.
Thus, the results presented here should 
be useful for identifying the ${\cal N}=2$ conformal field theory duals 
to the families of  AdS$_4\times Y^{p,k}_7$ backgrounds \cite{paper3}.

In the regime of parameters where $p^5 >> N >> p$, the backgrounds that we discuss are  better described
as type IIA solutions of the form AdS$_4\times M_6$, with non-trivial dilaton,   $F_4$ and $F_2$ RR fluxes.
The non-trivial dilaton comes from the norm of the Killing vector along which we reduce,
and  is naturally dictated by the construction of the metrics in \cite{paper3}.
The reduction preserves the ${\cal N}=2$ supersymmetries.
In particular, this is a different reduction to that considered in \cite{ABJM}.

The rest of the paper is organised as follows.  In section \ref{firstsection}
we review the Sasaki-Einstein metrics presented in \cite{paper3},
and for the cases of interest determine explicitly their dependence on the two integers $p$ and $k$. In section \ref{torics}
we compute the toric data of the $Y^{p,k}(\C P^2)$ and $Y^{p,k}(\C P^1 \times \C P^1)$ Calabi-Yau four-folds. We present the toric
diagrams and GLSM charges. In section \ref{homosection} we compute the homology of the manifolds and discuss supersymmetric
five-submanifolds. In section \ref{sugrasection} we discuss the AdS$_4\times Y^{p,k}_7$ M-theory backgrounds and their reduction to type
IIA supergravity. The limiting case AdS$_4\times Y^{p,3p} (\C P^2)$ is described in some detail. We conclude 
in section \ref{disc}.

%%%%%%%%%%%%%%%%%%%%%%%%%%%%%%%%%%%%%%%%%%%%%%%%%%%%%%%%%%%%%%%%%%%

\section{Metrics and volumes}
\label{firstsection}

\subsection{Review of the metrics of \cite{paper3}}
\label{revpaper3}

In this section we briefly recall the construction of the metrics of \cite{paper3}, and compute the volumes
of the corresponding manifolds. We initially keep the K\"ahler-Einstein manifold 
$(B_{2n},\tilde{g})$ general, specialising
to the two examples of interest, $B_4 = \C P^2$ and $B_4 = \C P^1 \times \C P^1$, only when it is necessary.

Take any complete $2n$-dimensional positive curvature
K\"ahler-Einstein manifold $B_{2n}$, with line element $\diff\tilde s^2$ and K\"ahler
form\footnote{Note that the one-form $A$ is only defined locally. In fact $A$ is a connection
on the anti-canonical line bundle of $B_{2n}$.}
$\tilde J =\diff A/2$.
The metric is normalised so that $\widetilde{\mathrm{Ric}} = \lambda \tilde{g}$.
Given any such $B_{2n}$, there is a countably infinite family of
associated Sasaki-Einstein metrics on the total space of certain
Lens space bundles
$S^3/\Z_p$ over $B_{2n}$.
The local metrics were presented in \cite{paper3} in the following form
\bea
   \diff s^2\, =\, \rho^2\diff\tilde s^2+U(\rho)^{-1}\diff\rho^2
      + q(\rho) (\diff\psi + A)^2
      + w(\rho) \left[\diff\alpha + f(\rho)(\diff\psi +A)\right]^2
\label{new}
\eea
where the function $U(\rho)$ is conveniently written as
\be
U(\rho)\,=\,\frac{\lambda}{2(n+1)(n+2)}\frac{1}{x^{n+1}}P(x;\kappa)~,
\qquad\quad
\,x=\,\frac{\Lambda}{\lambda}\rho^2
\ee
and
\bea
\label{polyx}
P(x;\kappa) \,= \, -(n+1)x^{n+2}+(n+2)x^{n+1}+\kappa~.
\eea
The remaining metric functions are then
\bea
   w(\rho) & = & \rho^2 U(\rho)+(\rho^2-\lambda/\Lambda)^2\nn\\[2mm]
   q(\rho) & = &
      \frac{\lambda^2}{\Lambda^2}\frac{\rho^2U(\rho)}{w(\rho)}\nn \\[2mm]
   f(\rho) & = &
      \frac{\rho^2(U(\rho)+\rho^2-\lambda/\Lambda)}{w(\rho)}~.
\label{deff}
\eea

In \cite{paper3} (see also \cite{resolutions}) it was shown that for
\bea \label{kappadude}
-1~<~\kappa ~< ~0
\eea
one can take the ranges of the coordinates $0\leq \psi \leq 4\pi/\lambda$
and $\rho_1\leq \rho \leq \rho_2$ so that the ``base'' $M_{2n+2}$
(excluding the $\alpha$ direction in (\ref{new}))
is the total space of an $S^2$ bundle over $B_{2n}$. In particular,
$\rho_i$ are the two positive roots of the equation $U(\rho)=0$ and satisfy
the inequalities
\bea
    0 ~<~ \rho_1 ~<~ \sqrt{\frac{\lambda}{\Lambda}}
     ~  <~ \rho_2 ~<~ \sqrt{\frac{\lambda (n+2)}{\Lambda(n+1)}}~.
\eea
The $S^2$ fibre is then coordinatised by the polar coordinate
$\rho$ and the axial coordinate $\psi$. Without loss of
generality we now set $\lambda=2$ and
$\Lambda=2(n+2)$,
so that the Sasaki-Einstein metric has Ricci curvature $2n+2$ times the metric.

For appropriate values of $\kappa$ in the above range,
one can periodically identify the $\alpha$ coordinate so as to
obtain a principle $U(1)$
bundle over the space $M_{2n+2}$.
Recall that the group $H_2(M_{2n+2};\Z)$ of two-cycles on $M_{2n+2}$ is
naturally $\Z\oplus H_2(B_{2n};\Z)$ where the
first factor is generated by a copy $\Sigma$ of the fibre $S^2$, and
the generators $\Sigma_i$ of $H_2(B_{2n};\Z)$
are pushed forward into $M_{2n+2}$ by the map
$\sigma^N:B_{2n}\rightarrow M_{2n+2}$, which denotes the section of
$\pi:M_{2n+2}\rightarrow B_{2n}$ corresponding to the ``north
pole" $\rho=\rho_2$ of the $S^2$ fibres.
One can then periodically identify $\alpha$ to obtain a principle
$U(1)$ bundle over $M_{2n+2}$ provided
 $B\equiv f(\rho)(\diff\psi+A)$ is proportional to a connection one-form.
This is true if and only if
the periods of $\tfrac{1}{2\pi}\diff B$ over the
representative basis $\{\Sigma,\sigma^N\Sigma_i\}$ are rationally related.
Equivalently, one
ensures that the periods of $\tfrac{\ell^{-1}}{2\pi}\diff B$ are all
integers, for some positive constant $\ell \in \R$.

The periods are easily computed\footnote{The definitions here
are slightly different to those in \cite{paper3}.}
to be
\bea
f(\rho_2)-f(\rho_1) & = & \int_{\Sigma} \frac{\diff B}{2\pi}~\equiv~\ell p\label{period1}\\
f(\rho_2) c_{(i)} & = & \int_{\sigma^N\Sigma_i} \frac{\diff B}{2\pi} ~\equiv~\ell
\frac{k}{h} c_{(i)}~,\label{period2}
\eea
where
\bea
c_{(i)}\,=\,\int_{\Sigma_i}\frac{\diff A}{2\pi}
  \, =\, \left<c_1({\cal L}),[\Sigma_i]\right>\in\Z
\eea
are Chern numbers of the anti-canonical 
bundle\footnote{Note, in particular, that for $B_4=\C P^2$, this is ${\cal L} = {\cal O}(3)_{\C P^2}$.} 
$\mathcal{L}$ over
$B_{2n}$ and we have defined $h=\mathrm{hcf}\{c_{(i)}\}$.
Thus we see that, if
$f(\rho_1)/f(\rho_2)$ is rational and hence $p,k\in\Z$, $\alpha$ can be
periodically identified with period $2\pi\ell$. The
$U(1)$ principle bundle, with coordinate $\gamma\ell\equiv
\alpha$, then has Chern numbers $\{p,kc_{(i)}/h\}$ with respect
to the basis $\{\Sigma,\sigma^N\Sigma_i\}$. The range of $k$ is fixed so that
\be
\frac{hp}{2}\,<\,k\,<\, h p\label{range}~,
\ee
as follows from the bound (\ref{kappadude}) on $\kappa$.

Note that the resulting Sasaki-Einstein manifold is indeed a Lens space bundle
$S^3/\Z_p$ over $B_{2n}$ -- this follows since the Chern number
of the $U(1)$ principle bundle over the fibre $S^2$ is $p$, which thus
forms a Lens space fibre. We will see later that these Chern numbers
may be re-interpreted as units of RR fluxes in a dual type IIA picture.
For further details on the construction of \cite{paper3}, see also \cite{resolutions}.

%%%%%%%%%%%%%%%%%%%%%%%%%%%%%%%%%%%%%%%%%%%%%%%%%%%%%%%%%%%%%%%%%%%%%%%

\subsection{Volumes}

We now turn to an analysis of the volumes of these manifolds.
First, it will be useful to note the following formulae
\bea
f(\rho_i) \,=\, \frac{\rho_i^2}{\rho_i^2-\tfrac{1}{n+2}}~,
\label{useless}
\eea
where $\rho_1, \rho_2$ are roots of the $(n+2)$-order polynomial.
One also easily derives the following relations
\bea
h\rho_1^2  \,=\,  (k-hp)\ell \left(\rho_1^2-\tfrac{1}{n+2}\right)~,\qquad \quad
h\rho_2^2  \,=\,  k\ell \left(\rho_2^2-\tfrac{1}{n+2}\right)~.
\eea
Defining $x_i=(n+2)\rho_i^2$, the integrated volume may be written as
\bea
\mathrm{vol} (Y_{2n+3}^{p,k}(B_{2n})) \, = \, \mathrm{vol} (B_{2n})
\frac{2\pi^2}{(n+1)(n+2)^{n+2}}\ell (x_2^{n+1}-x_1^{n+1})~
\label{volumefor}
\eea
where
\bea
\ell \, =\,  \frac{x_2-x_1}{p(x_2-1)(1-x_1)}~.
\label{alpharadius}
\eea
It is interesting to compute the formal limiting values of the volume
formula (\ref{volumefor})
in the limit that $k$ approaches the endpoints of
the interval (\ref{range}).
The case $k\to hp$ corresponds to $\kappa \to 0$.
It follows that $\ell \to \tfrac{n+2}{p}$, and the volume approaches
\bea
\vol (Y_{2n+3}^{p,k})\, \stackrel{k\to hp}{\longrightarrow}\, \vol (B_{2n}) \frac{2\pi^2}{p \,(n+1)^{n+2}}~.
\label{limit1}
\eea
The case $k\to hp/2$ corresponds to $\kappa \to -1$.
The limiting value of the volume is easily computed to be
\bea
\vol (Y_{2n+3}^{p,k})\, \stackrel{ k\to hp/2}{\longrightarrow} \, \vol (B_{2n}) \frac{8\pi^2}{p \, (n+2)^{n+2}}~.
\label{limit2}
\eea
Notice that in both cases the volumes are rational
multiples of the volume of the round sphere
$S^{2n+3}$.
For $n=1$, from (\ref{limit1}) and (\ref{limit2})
we correctly obtain\footnote{Note that the volume of the
K\"ahler-Einstein base $B_{2n}$ is normalised so that
$\widetilde{\mathrm{Ric}} = 2 \tilde{g}$. }
the values  $\tfrac{\pi^3}{2p}$
and  $\tfrac{16\pi^3}{27p}$, respectively \cite{toric}.
To be more explicit one should determine the roots $x_i$ in terms of the
integer parameters $p,k$. To this end, let us define the polynomial
\bea
Z (x_1,x_2) \, = \, \sum_{i=0}^n  x_1^i x_2^{n-i}~.
\eea
The defining equation of the roots $x_i$ is then generally
\bea
(n+1)Z_{n+1}(x_1,x_2) \, = \, (n+2)Z_{n}(x_1,x_2) ~.
\label{thecubic}
\eea
This is an $(n+1)$-th order
equation in the two variables $x_1,x_2$. To determine the roots we combine (\ref{thecubic})
with another relation that  may be obtained by  eliminating $\ell$ from the equations
(\ref{period1}),  (\ref{period2}) defining the periods. This yields
\bea
\frac{x_1(x_2-1)}{x_2(x_1-1)} \, = \, 1-\frac{hp}{k}~.
\eea
After solving for one of the roots and substituting back into (\ref{thecubic}), one obtains the final equation
from which the roots may be extracted. In the case $n=1$ one can check that
the quadratic equation in \cite{paper2} is reproduced.
 For our purposes, it suffices to analyse the case of $n=2$. %After some algebra
We obtain cubic equations defining the two roots:
\bea
&& \!\!\!\!\!\!3p^3\,x_1^3+2p^2(6b-5p)\,x_1^2+p(18b^2-28pb+11p^2)\,x_1+4(3b^3+4p^2b-6pb^2-p^3) \, = \, 0 \nn \\
&& \!\!\!\!\!\!3p^3\,x_2^3+2p^2(p-6b)\,x_2^2+p(18b^2-8pb+p^2)\,x_2 +4b(3pb -3b^2  -p^2)  \,= \, 0~,
\eea
where we have defined $b=k/h$.
These may be solved analytically, although the resulting expressions are lengthy. 
However, it is interesting to note that the
volumes are written in terms of cubic irrational numbers.

Note that for $B_{4}=\C P^2$, the first non-trivial example has $p=1$, $k=2$. 
In this case one easily computes
\bea
x_1&=&\frac{1}{9}\left[2+\left(53+6\sqrt{78}\right)^{1/3}+\frac{1}{\left(53+6\sqrt{78}\right)^{1/3}}\right]\ \approx \ 0.77\\
x_2&=&\frac{1}{9}\left[6+\left(27+3\sqrt{78}\right)^{1/3}+\frac{3}{\left(27+3\sqrt{78}\right)^{1/3}}\right]\ \approx \ 1.17~,
\eea
giving the volume formula
\bea
\mathrm{vol}\left(Y^{1,2}_7(\C P^2)\right)&=&\frac{3\pi^4}{64}\Bigg[\frac{107}{27} +\left(\frac{521}{54}-\sqrt{78}\right)\left(53+6\sqrt{78}\right)^{1/3} \nonumber\\ 
&&+\left(\frac{2341}{54}-\frac{44}{9}\sqrt{78}\right)\left(53+6\sqrt{78}\right)^{2/3}\Bigg]~.
\eea
It would be nice to reproduce these numbers from a field theory calculation.

\section{Toric description}
\label{torics}

Provided the base K\"ahler-Einstein manifold $(B_{2n},\tilde{g})$ is toric,
the Calabi-Yau cones in two complex dimensions higher
are also toric. One can analyse these explicitly following the
techniques described in \cite{toric} for the case of $n=1$.
The general idea is simple. The Calabi-Yau cones
\be
\diff s^2 (\mathrm{CY}_{2n+4}) \,=\, \diff r^2 + r^2 \diff s^2 (Y_{2n+3})
\ee
have a Hamiltonian torus action by $\T^{n+2}$, and so by definition are toric.
Here the K\"ahler form $\omega$ of the Calabi-Yau
may be regarded as a symplectic form, and one can then introduce a
moment map
$\mu:C(Y_{2n+3})\rightarrow \R^{n+2}$.
The image is always a convex rational polyhedral cone, of a
special type, and the moment map exhibits the Calabi-Yau as a $\T^{n+2}$
fibration over this polyhedral cone. Writing the symplectic form of
$B_{2n}$ as
\be
\tilde J \,= \, \diff \phi_i \wedge \diff \mu^i_{B_{2n}}~,
\ee
the symplectic form of the Calabi-Yau cones may be written as
\bea
\omega\, =\, \diff \phi_i \wedge \diff \Big[ r^2\rho^2 \mu^i_{B_{2n}}\Big] + \diff \psi \wedge \diff \Big[ -\tfrac{1}{2}r^2\rho^2\Big]+ \diff \gamma \wedge \Big[  \tfrac{\ell}{2}r^2(\tfrac{1}{n+2}-\rho^2)\Big]~.
\label{little}
\eea
From this it is fairly immediate to read off the moment map.
However, a remaining problem is to determine a choice of angular coordinates,
and correspondingly the choice of moment map coordinates,
such that the associated vector fields generate
an effectively acting $\T^{n+2}$. This coordinate basis will be unique
up to $SL(4;\Z)$.
In the remainder of this section
we compute the toric and linear sigma model
descriptions of $Y^{p,k}(\C P^2)$ and $Y^{p,k}(\C P^1\times \C P^1)$ using the
techniques described in \cite{toric}, to which we refer for further details.
For the time being we assume $\mathrm{hcf}(p,k)=1$.

\subsection{$Y^{p,k}(\C P^2)$ family}

Recall that $\C P^2$ equipped with its Fubini-Study metric
is a toric K\"ahler-Einstein manifold. In terms of
homogeneous coordinates the torus
action is
\be
[z_0,z_1,z_2]\rightarrow
[z_0,\exp(i\phi_1)z_1,\exp(i\phi_2)z_2]\ee which has moment map
$\mu_{FS}:\C P^2 \rightarrow \R^2$ given by
\be
\mu_{FS} = -\frac{3}{2}\left(\frac{|z_1|^2}{|z_0|^2+|z_1|^2+|z_2|^2},
\frac{|z_2|^2}{|z_0|^2+|z_1|^2+|z_2|^2}\right)~.\ee
Here we have normalised the metric so that $\mathrm{Ric}=2g_{FS}$.
As is well-known, the image in $\R^2$ is a triangle with vertices
$(0,0)$, $(-3/2,0)$, $(0,-3/2)$. The canonical bundle over
$\C P^2$ has Chern class $-3$, and hence $h=3$.  Note we may take
\bea
A \,=\,  -2\mu_{FS}^i\diff\phi_i~.
\eea

The following is a basis for an effectively
acting $\T^4$ on $Y^{p,k}(\C P^2)$:
\bea\label{basiscp2}
e_1  =  \frac{\partial}{\partial\phi_1}-\frac{\partial}{\partial\psi}
+ \frac{k}{3}
\frac{\partial}{\partial\gamma}, \quad
e_2  =  \frac{\partial}{\partial\phi_2}-\frac{\partial}{\partial\psi}
+ \frac{k}{3}
\frac{\partial}{\partial\gamma},\quad
e_3  = \frac{\partial}{\partial\psi}-\frac{k}{3}
\frac{\partial}{\partial\gamma},\quad
e_4  = \frac{\partial}{\partial\gamma}~.
\eea
The appearance of the fractional terms $k/3$ is crucial in order
that the orbits of the group action close, giving an effective action
of the torus on the Calabi-Yau cone. This issue
was discussed in \cite{toric}, and there is a straightforward way to
fix a good basis of angular coordinates. Consider, for example, the fixed
complex ray $\C^*$ given by $\{z_1=z_2=0, \rho=\rho_2\}$. 
The induced metric is
\bea
\diff r^2 + r^2 \ell^2 w(\rho_2) \left(\diff \gamma + \frac{k}{3}\diff\psi\right)^2~.
\eea
Thus we define the new coordinates
\bea
\phi_3=\psi, \qquad
\phi_4 = \gamma+\frac{k}{3}\psi~,
\eea
and note that $\phi_4$ is a periodic coordinate on the $\C^*$, and that
$\phi_1$, $\phi_2$ and $\phi_3$ coordinatise the $\T^3$ that fixes
this line. Thus $\phi_a$, $a=1,\ldots,4$, may be taken to be standard
coordinates on $\T^4$. Note that
\bea
\frac{\partial}{\partial \phi_3} = \frac{\partial}{\partial\psi}-\frac{k}{3}\frac{\partial}{\partial \gamma}, \qquad
\frac{\partial}{\partial\phi_4} = \frac{\partial}{\partial \gamma}~.
\eea
This basis is unique only up to $SL(4;\Z)$
transformations. For example, one easily checks that the natural
induced bases at the other rays are equivalent to the one above.
In (\ref{basiscp2}) we have chosen a slightly different, but particularly convenient, basis.
The moment map in this basis is then
\bea
\mu & = & r^2\Big[\rho^2\mu_{FS}^1 + \tfrac{1}{2}\rho^2 -\tfrac{1}{6}k\ell
\left(\rho^2-\tfrac{1}{4}\right) , \rho^2\mu_{FS}^2 + \tfrac{1}{2}\rho^2 -\tfrac{1}{6}k\ell
\left(\rho^2-\tfrac{1}{4}\right) ,  \nn \\
& & -\tfrac{1}{2}\rho^2+ \tfrac{1}{6}k\ell
\left(\rho^2-\tfrac{1}{4}\right) ,
 -\tfrac{1}{2}\ell \left(\rho^2-\tfrac{1}{4}\right)\Big]~.\eea
This is easily computed using the K\"ahler form (\ref{little}) of the
Calabi-Yau cone.

We now identify the half-lines which form the polyhedral cone. These are
submanifolds of $Y_7$ over which a $\T^3$ collapses. They
are precisely the collection of 6 circles given by the vanishing of
all but one of the 3 homogeneous coordinates on $\C P^2$, together with
 $\rho=\rho_1,\rho_2$.
Noting that $\rho_1^2-\tfrac{1}{n+2}<0$ and $\rho_2^2-\tfrac{1}{n+2}>0$,
these half-lines are spanned by the vectors in $\R^4$:
\begin{equation}
\begin{aligned}
   u_1  = [p,p,-p,1]~, &   \quad u_2 =[-2p+k,p,-p,1]~, & \quad u_3 & =[p,-2p+k,-p,1]~, \\
    u_4  = [0,0,0,-1]~, &  \quad  u_5 =[-k,0,0,-1]~,       &\quad  u_6 & =[0,-k,0,-1] ~,
\end{aligned}
\label{looks}
\end{equation}
where the first 3 vectors correspond to $\rho=\rho_1$ and the
remaining 3 correspond to $\rho=\rho_2$.
These vectors form a convex rational polyhedral cone, and it is
simple to compute the outward pointing primitive
normal vectors to the facets of this cone.
There are 5 facets with normal vectors
\be
v_1  =  [0,0,1,0], ~v_2 = [0,0,1,p], ~v_3 = [1,0,1,0], ~
v_4 = [0,1,1,0], ~v_5  =  [-1,-1,1,k]~.
\label{cpvectors}
\ee
As in \cite{toric}, it will be useful to obtain a 
gauged linear sigma model description of the geometry. 
Here,  in order to keep the paper relatively 
self-contained, we give a lightning review of gauged linear sigma 
models and Delzant's theorem \cite{L}, referring to 
\cite{toric} for further details. 
Let $z_1,\ldots,z_d$ denote complex coordinates on $\C^d$. In physics terms, these will be the
lowest components of chiral superfields $\Phi_i$, $i=1,\ldots,d$. We may specify an action of the
group $\T^r\cong U(1)^r$ on $\C^d$ by giving the integral charge matrix $Q=\{Q^i_{a}\mid i=1,\ldots,d ; \ a=1,\ldots,r\}$; here
the $a$th copy of $U(1)$ acts on $\C^d$ as
\be
(z_1,\ldots,z_d)\,\rightarrow\,  (\lambda^{Q^1_a}z_1,\ldots,\lambda^{Q^d_{a}}z_d)\ee
where $\lambda\in U(1)$. We may then
perform the so-called K\"ahler quotient $X=\C^d//U(1)^r$ by imposing the $r$ constraints
\be
\sum_{i=1}^d Q^i_a |z_i|^2 \,=\, t_a \quad\qquad a\,=\,1,\ldots,r~,
\label{Dterms}
\ee
where $t_a$ are constants, and then quotienting by $U(1)^r$. The resulting space $X$ has complex dimension
$m=d-r$ and inherits a K\"ahler, and hence also symplectic, structure from that of $\C^d$. In physics terms,
the constraints (\ref{Dterms})
correspond to setting the $D$-terms of the gauged linear sigma model to zero to give the vacuum, 
where $t_a$ are
FI parameters. The quotient by $\T^r$ then removes the gauge degrees of freedom. Thus the K\"ahler quotient
of the gauged linear sigma model precisely describes the classical vacuum of the theory. 
For the cases of interest in this paper, we set $t_a=0$ so that the resulting quotient space
is a cone. It is also an important fact that $c_1(X)=0$ is equivalent to the 
statement that the sum of the $U(1)$ charges
is zero for each $U(1)$ factor. Thus
\be
\sum_{i=1}^d Q^i_a \,= \, 0\quad\quad a=1,\ldots,r~.
\ee
The sigma model is then Calabi-Yau, although
note that the metric induced by the K\"ahler quotient is \emph{not} in general Ricci-flat.

In order to go from the moment map description to the gauged linear sigma model 
description above, one can apply the 
Delzant theorem of \cite{L}. We begin by considering the linear map $\pi: \R^d \rightarrow \R^m$ 
which maps the standard basis
 vectors $E_i$ of $\R^d$
to the outward normal vectors $v_i$ of the moment polytope. 
Thus $\pi(E_i)=v_i$ for each $i=1,\ldots,d$. Moreover, since the map maps lattice vectors to 
lattice vectors, one also obtains an induced map of tori
\bea
\tilde{\pi}: \, \T^d\rightarrow \T^m~.
\eea
The Delzant theorem is that the gauged linear sigma model gauge group is the kernel of the 
map $\tilde{\pi}$. Note this may contain discrete factors, so that the kernel is 
not connected. This will occur, for example, for the orbifold $\C^4/\Z_{3p}$ discussed below.

In the case at hand, we must compute the kernel of the map
\bea
\R^5\rightarrow \R^4: \quad E_a\mapsto v_a~.
\eea
Thus $d=5$, $m=4$, in the above notation. The kernel is generated by the primitive vector in the
integral lattice $\Z^5$ given by
\bea
(-3p+k,-k,p,p,p)~.
\eea
These are thus the charges of the gauged linear sigma model. 

\begin{figure}[ht!]
 \epsfxsize = 5cm
\centerline{\epsfbox{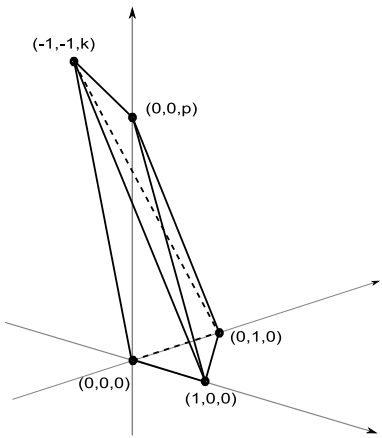}}
\caption{Toric diagram for $Y^{p,k}(\C P^2)$. The polytope is bounded by 6 
triangular faces.}\label{toricvectors1}
\end{figure}

Note that the vectors (\ref{cpvectors}) are coplanar, all lying
on the plane $\{E_3=1\}$. This is a result of the Calabi-Yau condition.
We may hence represent the toric data as a set of vectors in
$\Z^3$:
\be
w_1  =  [0,0,0], \quad w_2 = [0,0,p], \quad w_3 = [1,0,0], \quad
w_4 = [0,1,0], \quad  w_5  =  [-1,-1,k]~.\label{5legs}
\ee
It is a general result that these vectors form the vertices 
of a compact convex lattice polytope in $\Z^3\subset\R^3$. The 
corresponding diagram is usually called the toric diagram in the 
physics literature. The polytope for $Y^{p,k}(\C P^2)$ is shown in Figure \ref{toricvectors1}.

\begin{figure}[ht]
  \begin{minipage}[t]{0.48\textwidth}
    \begin{center}
    \epsfxsize=5cm
    \epsfbox{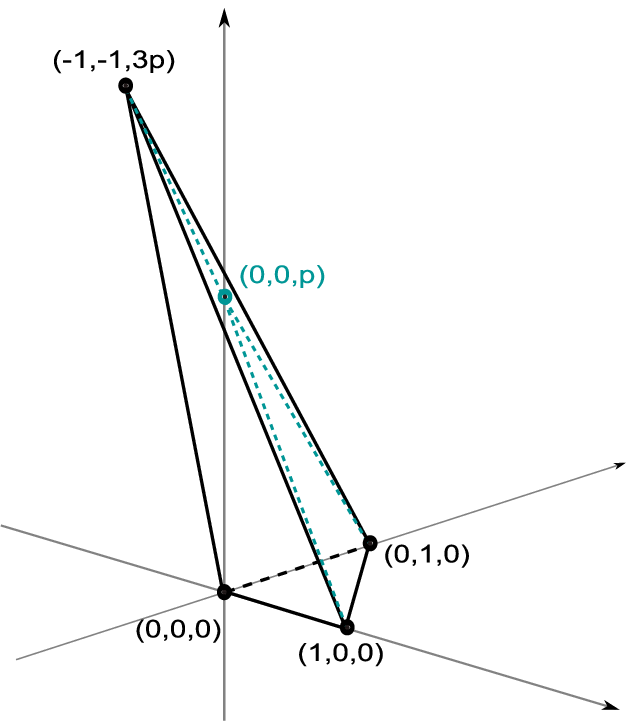}
     %\caption{}
      \end{center}
  \end{minipage}
  \hfill
  \begin{minipage}[t]{.48\textwidth}
    \begin{center}
     \epsfxsize=5cm
    \epsfbox{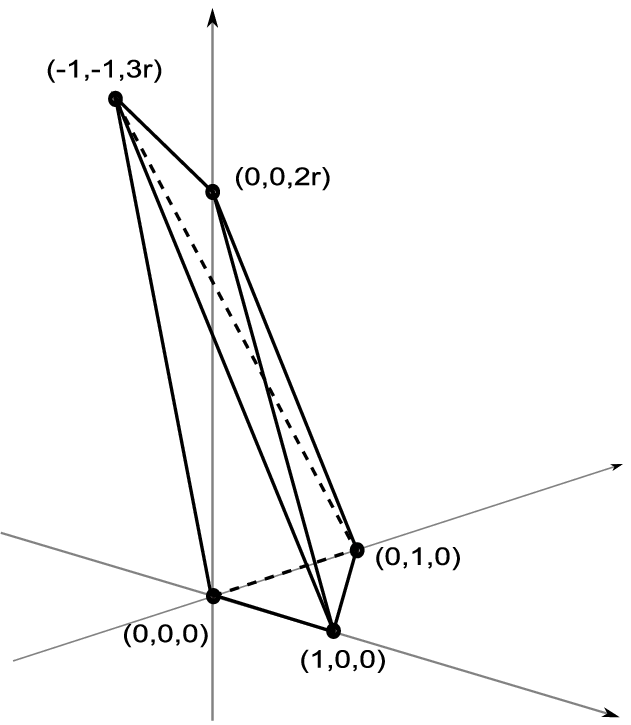}
     % \caption{}
     \end{center}
  \end{minipage}
\caption{On the left hand side: Toric diagram for the orbifold $Y^{p,3p}(\C P^2)=\C^4/\Z_{3p}$. The polytope 
is bounded by 4 triangular faces. On the right hand side: Toric diagram for $Y^{2r,3r}(\C P^2)=M^{3,2}/\Z_r$.
}
  \hfill
\label{toricorb}
\end{figure}

This information allows for a simple identification of the limits $k=3p$ and
$2k=3p$. In the former case one can show that the vector
$w_2$ lies on the plane defined by $w_3,w_4,w_5$. We may hence
discard this vector to give a minimal presentation of the
singularity. Thus this limit is necessarily an orbifold
of $\C^4$. Using the Delzant theorem one easily finds the
orbifold action is generated by
\be
(\omega_{3p}, \ \omega_{3p}, \ \omega_{3p}, \ \omega_{3p}^{-3})\in SU(4)
\label{orbi3p}
\ee
where  $\omega_{3p}$ is a $3p$-th root of unity. We thus
obtain the orbifold $\C^4/\Z_{3p}=(\C^4/\Z_3)/\Z_p$. The toric diagram 
is shown in Figure \ref{toricorb}.  We shall 
return to consider this orbifold in more detail later.

The limit $2k=3p$ clearly requires $p=2r$ even. One then has
a $\Z_r$ orbifold of the gauged linear sigma model with charges
\be
(-3,-3, 2,2,2)~.
\ee
In fact this is just the complex cone over $\C P^2\times \C P^1$.
The Sasaki-Einstein metric is the homogeneous metric known as
$M^{3,2}$.
The finite quotient is given by $\Z_r\subset SU(2)$ which
acts on the $\C P^1$, thus breaking the isometry group to
$SU(3)\times U(1)^2$ which is the isometry group of the general
$Y^{p,k}(\C P^2)$ manifold. The toric diagram is shown in Figure \ref{toricorb}.
Equations\footnote{Recall that, in our normalisation for the
 K\"ahler-Einstein base, $\vol(\C P^2)=9\pi^2/2$.} (\ref{limit1}) and (\ref{limit2})
 show that  the volume of a generic $Y^{p,k}(\C P^2)$ lies within the following range
\be
\frac{9\pi^4}{128r}= \vol (M^{3,2}/\Z_r) > \vol (Y^{p,k} (\C P^2)) > \vol (S^7/\Z_{3p}) =
\frac{\pi^4}{9p}
\label{rangevolh3}
\ee
where the volume\footnote{This volume was also computed in \cite{italians}.}
of $M^{3,2}$ is easily computed using the topological formula
\bea
\vol (M^{3,2}) & = & \frac{\pi^4}{768} \int_{\C P^2 \times \C P^1} c_1^3~.
\label{topform}
\eea
It is interesting to notice that, at fixed $p$, the volume is a monotonically decreasing function of $k$
in the range (\ref{rangevolh3}).

%%%%%%%%%%%%%%%%%%%%%%%%%%%%%%%%%%%%%%%%%%%%%%%%%%%%%%%%%%%%%%%%%%%%%%%%%%%%

\subsection{$Y^{p,k}(\C P^1\times \C P^1)$ family}

Since the canonical bundle over $\C P^1\times \C P^1$ has
both Chern numbers equal to $-2$, we have $h=2$.
The following is a basis for an effectively acting $\T^4$ on $Y^{p,k}(\C P^1
\times \C P^2)$:
\bea\label{basisS2}
e_1 & = & \frac{\partial}{\partial\phi_1}-\frac{k}{2}
\frac{\partial}{\partial\gamma}, \quad
e_2  = \frac{\partial}{\partial\phi_2}-\frac{k}{2}
\frac{\partial}{\partial\gamma},\quad
e_3  = \frac{\partial}{\partial\psi}-\frac{k}{2}
\frac{\partial}{\partial\gamma},\quad
e_4  = \frac{\partial}{\partial\gamma}~.
\eea
Here $\phi_j$, $j=1,2$, are azimuthal coordinates on the two copies
of $\C P^1$, respectively. The argument that leads to the basis (\ref{basisS2})
is similar to that in the previous subsection.
The moment map in this basis is
\bea
\mu & = & r^2\Big[\tfrac{1}{2}\rho^2\cos\theta_1+\tfrac{1}{4}k\ell
\left(\rho^2-\tfrac{1}{4}\right), \tfrac{1}{2}\rho^2\cos\theta_2+\tfrac{1}{4}k\ell
\left(\rho^2-\tfrac{1}{4}\right), \nn \\
& & -\tfrac{1}{2}\rho^2+ \tfrac{1}{4}k\ell
\left(\rho^2-\tfrac{1}{4}\right),
 -\tfrac{1}{2}\ell \left(\rho^2-\tfrac{1}{4}\right)\Big]~.\eea
Here $\theta_1, \theta_2$ are usual polar coordinates on the two
two-spheres.

We now identify the half-lines which form the polyhedral cone. These
are precisely the collection of 8 circles given by all $2^3$ combinations of
 $\theta_1=0,\pi$; $\theta_2=0,\pi$; $\rho=\rho_1,\rho_2$.
These half-lines are spanned by the following vectors in $\R^4$:
\be
\begin{aligned}
& u_1 = [-k+p,-k+p,-p,1], & \quad &  u_2=[-k+p,-p,-p,1], &\\
& u_3 = [-p,-k+p,-p,1], & \quad &  u_4 = [-p,-p,-p,1], &\quad & u_5 = [k,k,0,-1],\nn \\
& u_6=[k,0,0,-1],      &\quad &  u_7=[0,k,0,-1],& \quad  & u_8 =  [0,0,0,-1]~,
\end{aligned}\ee
where the first 4 vectors correspond to $\rho=\rho_1$ and the
remaining 4 correspond to $\rho=\rho_2$.
There are 6 facets for this polyhedral cone with normal vectors
\be
\begin{aligned}
& v_1  = [0,0,1,0], &\quad &v_2 = [0,0,1,p], & \quad & v_3  =  [-1,0,1,0],  \\
& v_4 = [1,0,1,k], & \quad & v_5 = [0,-1,1,0],&  \quad &   v_6 = [0,1,1,k]~.
\label{qwe}
\end{aligned}
\ee
Each of these vectors has zero dot products with precisely four of the $u_i$
and has negative dot products with the remaining four.

\begin{figure}[ht!]
 \epsfxsize = 5cm
\centerline{\epsfbox{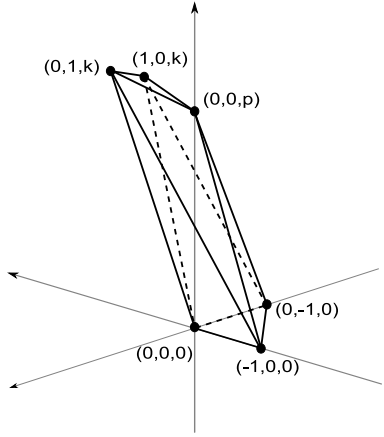}}
\caption{Toric diagram for $Y^{p,k}(\C P^1\times\C P^1)$. The polytope is bounded 
by 8 triangular faces.}\label{toricvectors2}
\end{figure}

We may now apply the Delzant theorem of \cite{L}. Thus we compute the
kernel of the map
\bea
\R^6\rightarrow \R^4: \quad E_a\mapsto v_a\eea
where $E_a$, $a=1,\ldots,6$ is the standard orthonormal basis for
$\R^6$. This kernel is generated by the primitive vectors in the lattice
$\Z^6$
\bea
(-2p+k,-k, p,p,0,0)\nn \\
(-2p+k,-k, 0,0,p,p)\eea
which give the charges of the gauged linear sigma model.
Again note that the vectors (\ref{qwe}) lie on the plane $\{E_3=1\}$ and thus
we may project onto this plane to obtain
\be
\begin{aligned}
& w_1  =  [0,0,0], & \quad & w_2 = [0,0,p], & \quad & w_3 = [-1,0,0], \\
& w_4 = [1,0,k],  & \quad &  w_5  =  [0,-1,0], & \quad &  w_6 = [0,1,k]~.
\end{aligned}
\ee
The corresponding toric diagram for $Y^{p,k}(\C P^1 \times \C P^1)$ is shown in 
Figure \ref{toricvectors2}.

\begin{figure}[ht]
  \begin{minipage}[t]{0.48\textwidth}
    \begin{center}
    \epsfxsize=5cm
    \epsfbox{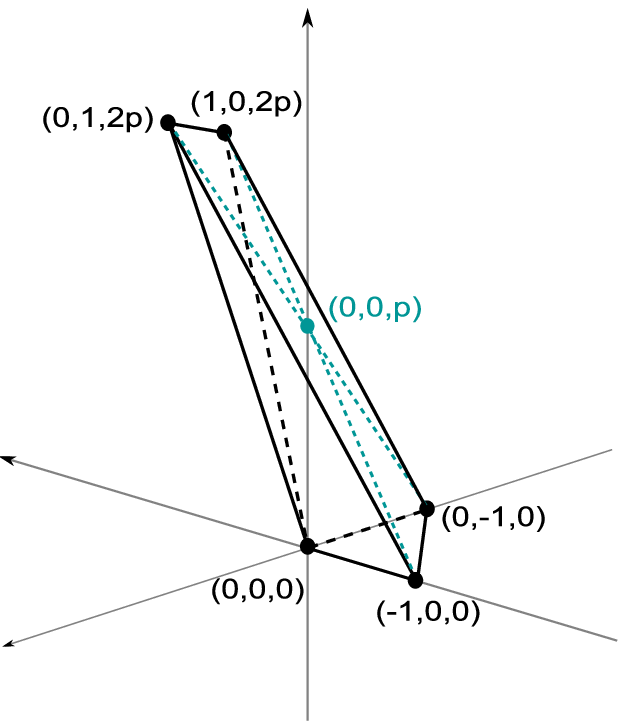}
     %\caption{}
      \end{center}
  \end{minipage}
  \hfill
  \begin{minipage}[t]{.48\textwidth}
    \begin{center}
     \epsfxsize=5cm
    \epsfbox{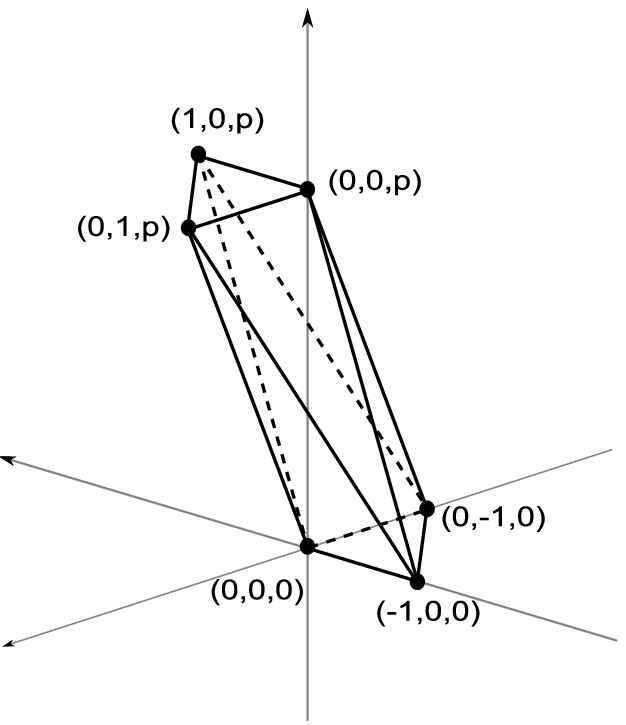}
     % \caption{}
     \end{center}
  \end{minipage}
\caption{On the left hand side: Toric diagram for $Y^{p,2p}(\C P^1\times\C P^1)$. This is bounded by 4 
triangles and a parallelogram, implying that the link of the singularity 
has worse-than-orbifold singularities.
 On the right hand side: Toric diagram for $Y^{p,p}(\C P^1\times\C P^1)=Q^{1,1,1}/\Z_p$.}
\hfill
\label{toricsingsvg}
\end{figure}

We now identify the limits $k=2p$, $k=p$. The former is a $\Z_p$
quotient of the gauged linear sigma model with charges
\bea
&& (0,-2,1,1,0,0) \nn\\
&& (0,-2,0,0,1,1)~.
\eea
In fact this describes $\C\times \mathcal{C}_{\C}(\C P^1\times\C P^1)$,
where $\mathcal{C}_{\C}(\C P^1\times\C P^1)$ denotes the complex
cone over $\C P^1\times\C P^1$. Thus the boundary of this space
has worse-than-orbifold singularities\footnote{The complex
cone over $\C P^2$ is an orbifold, which is why projective
spaces are exceptional in this limit.}.  One can also see this
from the toric diagram, shown in Figure \ref{toricsingsvg}. For general $p$ and $k$, the vertices $w_3$, $w_5$, $w_4$, $w_6$ form 
a parallelogram with edge vectors $(1,1,k)$ and $(1,-1,0)$. When $k=2p$,
the vertex $w_2=(0,0,p)$ lies in this parallelogram. Thus the parallelogram 
itself becomes a bounding face of the polytope; the fact that this 
is not a triangle implies that one has worse-than-orbifold singularities 
on the link of the singularity at the apex of the cone.

The limit $k=p$ is instead a $\Z_p$ quotient of the gauged linear sigma model with charges
\bea
&& (-1,-1,1,1,0,0)\nn\\
&&(-1,-1,0,0,1,1)~.
\eea
This space is the circle bundle over $\C P^1\times\C P^1\times\C P^1$
with Chern numbers 1 over each $\C P^1$, and the
corresponding homogeneous Sasaki-Einstein manifold is known as $Q^{1,1,1}$.
The finite quotient is given by $\Z_p\subset SU(2)$ in the first $\C P^1$
which thus breaks the isometry group to
$SU(2)\times SU(2)\times U(1)^2$. This is the isometry group of the general
$Y^{p,k}(\C P^1\times \C P^1)$ manifold. The toric diagram is shown in Figure \ref{toricsingsvg}.
Using equations (\ref{limit1}) and (\ref{limit2}), we find that the volume
of $Y^{p,k}(\C P^1\times \C P^1)$ lies within the following range
\be
\frac{\pi^4}{8p}= \vol (Q^{1,1,1}/\Z_p) > \vol (Y^{p,k} (\C P^1\times \C P^1)) > \vol (\partial L /\Z_p) =
\frac{8\pi^4}{81p}~,
\label{rangevolh2}
\ee
where the volume of $Q^{1,1,1}$ is easily computed using a topological formula similar to (\ref{topform}), and is also given
for instance in \cite{italians}. Again, at fixed $p$, the volume is a monotonically decreasing function of $k$
in the range (\ref{rangevolh2}).

%%%%%%%%%%%%%%%%%%%%%%%%%%%%%%%%%%%%%%%%%%%%%%%%%%%%%%%%%%%%%%%%%%%%%%%%%%%%%%%%%%%%%%%%

\section{Homology and supersymmetric submanifolds}
\label{homosection}

\subsection{Homology}

In this subsection we make some comments on the homology
of $Y=Y^{p,k}(B_{2n})$. We begin by keeping $B_{2n}$ general, specialising 
to the two cases of interest only when it is necessary.

Since for $(p,k)\equiv\mathrm{hcf}(p,k)=1$ the Chern
numbers of the $\alpha$ circle bundle are relatively prime,
it follows that $Y$ is simply-connected. More generally we have
$\pi_1(Y^{p,k})\cong \Z_{(p,k)}$.
Using the Gysin sequence for the circle fibration, it is
also easy to see that $H_2(Y)\cong\Z^{b_2(B_{2n})}$ where
$b_2(B_{2n})$ is the second Betti number of $B_{2n}$. In fact these
topological invariants are also easily deduced from the
toric data \cite{Ltop}. Specifically,
\bea
\pi_1(Y)  \,\cong\, \Z^{n+2}/<v_a> ~, \qquad \quad
\pi_2(Y)\,\cong \,\Z^{d-(n+2)}~,
\eea
where $d$ is the number of normals $\{v_a\}$.
Thus, in particular, the number of gauge groups in the gauged linear sigma
model is always given by $b_2(B_{2n})$.

The remaining homology groups are easily computed using the
Gysin sequence of the $\alpha$ circle bundle. The result for the
two cases studied in this paper is
\begin{equation}
\begin{aligned}
H_0\,\cong\,\Z, & \quad H_1\,\cong \,\Z_{(p,k)}, &\quad &H_2\cong \Z^{b_2(B_{2n})},& \quad &
H_3\,\cong \,\Gamma, & \\
H_4\,\cong \, 0,& \quad H_5 \,\cong \,\Z^{b_2(B_{2n})}\oplus \Z_{(p,k)}, & \quad&
H_6\, \cong \, 0,  & \quad & H_7\,\cong\,\Z~.
\end{aligned}
\end{equation}
Here the finite group $\Gamma$ is
\be
\Gamma \, \cong \left\{
\begin{array}{l}
 \Z^2/<(0,-3p+k),(k,p)>\\
\Z^3/<(0,-2p+k,-2p+k),(k,p,0),(k,0,p)>\end{array}\right.
\ee
in the case $B_4=\C P^2$ and $B_4=\C P^1\times\C P^1$, respectively.
To derive these last results it is useful to note that the
cohomology ring of $M_6$ is given by the polynomial ring
\be
H^*(M_6)\,\cong \, H^*(B_4)[z]/(z^2-c_1(\mathcal{L})z)
\ee
where $z$ generates the cohomology of the fibre $S^2$. This
follows since, topologically, $M_6$ is the projectivisation of
the bundle $\mathcal{O}\oplus\mathcal{L}\rightarrow B_{2n}$. The
cohomology ring of $M_6$ is then standard -- see \cite{bott}.
Then the Gysin sequence gives that
\be
H^4(Y_7)\,\cong\, H^4(M_6)/[c_1\cup H^2(M_6)]
\ee
where $c_1=pz+(k/h)\pi^*c_1(\mathcal{L})$ is the first
Chern class of the $\alpha$ circle bundle.

We conclude by summarising the non-zero Betti numbers for the two cases of interest:
\begin{equation}
\begin{array}{lll}
Y^{p,k}(\C P^2)~: \quad  & b_0 \, =\, b_7 \, =\, 1~,\qquad  & b_2 \, = \, b_5 \, =\, 1~.\\[2mm]
Y^{p,k}(\C P^1 \times \C P^1)~: \quad & b_0 \, =\, b_7 \, =\, 1~,\qquad  & b_2 \, = \, b_5 \, =\, 2~.
\end{array}
\end{equation}
This implies that in the dual gauge theories one expects to find 
one or two  global  ``baryonic'' $U(1)$ symmetries, respectively\footnote{%We do not really know how
%global symmetries of the three-dimensional field theories are realised geometrically.
Notice that, 
although $S^7$ has no five-cycles, the ABJM quiver theory has a global ``baryonic'' symmetry.}. These are
 associated
to massless gauge fields in AdS$_4$, coming from Kaluza-Klein reduction of
 the M-theory six-form (dual to the three-form) on the internal five-cycles. In fact, the values above 
are also valid for
the limiting cases $Y^{2,3}(\C P^2) =M^{3,2}$ and $Y^{1,1}(\C P^1 \times \C P^1)=Q^{1,1,1}$ \cite{italians}.

%%%%%%%%%%%%%%%%%%%%%%%%%%%%%%%%%%%%%%%%%%%%%%%%%%%%%%%%%%%%%%%%%%%%%%%%

\subsection{Supersymmetric submanifolds}

We now discuss supersymmetric
5-submanifolds.  By definition, one may wrap M5-branes on these submanifolds and preserve supersymmetry. These should correspond
to BPS baryon-like operators in the dual SCFT$_3$. In particular, the conformal dimensions (and R-charges)
of these operators are proportional to the corresponding volumes
of the submanifolds, and provide important checks on the conjectured dual field theories. Specifically, the conformal
dimension of such operators is given by \cite{GubserKlebanov}
\bea
\Delta \,=\, \frac{\pi N}{6} \frac{\vol(\Sigma_5)}{\vol(Y_7)}~,
\label{condim}
\eea
where $N$ denotes the  number of M2 branes, 
and should also be related to the rank of the gauge group in the dual CFT$_3$.

These submanifolds are the bases of six-dimensional cones which are divisors in the Calabi-Yau.
The toric divisors are the inverse images under the moment map of the
facets of the polyhedral cone. However, here we will characterise the submaniolds
using specific features of the construction of \cite{paper3}.
As we reviewed in section \ref{firstsection},
all Sasaki-Einstein manifolds constructed in \cite{paper3} arise as principle $U(1)_\alpha$ bundles
over certain manifolds $M_{2n+2}$, which are themselves $S^2$ bundles over K\"ahler-Einstein manifolds $B_{2n}$.
It is easy to show (see \cite{toric}) that taking a
section $\{\rho=\rho_i\}$ of the $S^2$ fibre and fibering
with $U(1)_\alpha$ gives rise to two supersymmetric
$(2n+1)$-submanifolds $\Xi_1,\Xi_2$. The volumes of these are given by
\bea
\vol (\Xi_{i}) \, = \, \vol (B_{2n}) 2\pi \ell \frac{x_i^{n}}{(n+2)^{n+1}} |x_i-1|\qquad i=1,2~.
\label{volxi}
\eea
For the $n=2$ cases discussed in this paper it is also easy to determine
their topology:
\bea
Y^{p,k}(\C P^2)&:&~ \left\{\begin{array}{l}
\Xi_1 \cong  S^5 /\Z_{3p-k} \\
\Xi_2 \cong S^5/\Z_k
\end{array}\right.
\quad  \\
Y^{p,k}(\C P^1 \times \C P^1)&:&~ \left\{\begin{array}{l}
\Xi_1 \cong  (S^2\times S^3)/\Z_{2p-k} \\
\Xi_2 \cong  (S^2\times S^3) /\Z_k
\end{array}\right.
\eea
The finite quotients are along the fibres of the principle circle bundles
$S^1\hookrightarrow S^5\rightarrow\C P^2$,
$S^1\hookrightarrow T^{1,1}\rightarrow\C P^1\times \C P^1$, respectively.

Now, if $B_{2n}$ is toric it will admit a number
of $(2n-2)$-dimensional
toric divisors $\{\sigma_i, i=1,\dots g\}$. These lift to non-compact
toric divisors on the Calabi-Yau $(n+2)$-fold whose boundaries
are $g$ additional supersymmetric $(2n+1)$-submanifolds $\Theta_i$
of $Y_{2n+3}$. Their volumes are given by
\bea
\vol (\Theta_i) \, = \, \vol (\sigma_i) \frac{2\pi^2}{n(n+2)^{n+1}}\ell (x_2^{n}-x_1^{n})~, \qquad  i =1,\dots, g~.
\eea

For $B_4=\C P^1 \times \C P^1$, notice that
these are in fact topologically four copies of $Y_5^{p,q}$, where $k=p+q$.
For $B_4=\C P^2$, the projection to $M_6$ gives topologically three copies of the
third Hirzebruch surface $\F_3$ -- that is, a $\C P^1$ bundle over $\C P^1$
with twist 3. In fact this space is diffeomorphic to $\F_1$. $\F_3$ is not a spin manifold and, in fact,
depending on $p$ and $k$, neither is the total space of the
$\alpha$ circle bundle over this. Thus, in these cases, these
supersymmetric submanifolds are not spin. However, note that both D-branes 
and M5-branes may still be wrapped supersymmetrically on non-spin manifolds\footnote{Although 
this may introduce additional subtleties. For example, the Freed-Witten anomaly 
shifts the periods of the world-volume gauge field to half-integer values on a non-spin manifold.}.
For reference, we write down the volumes
\begin{equation}
\begin{array}{lccll}
Y^{p,k}(\C P^2)~:                          & \quad\vol (\Theta_i) & = & \frac{3\pi^3\ell}{4^3}(x_2^2-x_1^2) &\quad i=1,2,3\\[2mm]
Y^{p,k}(\C P^1 \times \C P^1)~: & \quad \vol (\Theta_i) & =  &  \frac{2\pi^3\ell}{4^3}(x_2^2-x_1^2) &\quad i = 1,\dots,4~.
\end{array}
\end{equation}

In general, the $x_i$ are cubic roots and the expressions for these volumes are rather lengthy. However, it may be useful
to record the values in the orbifold limits. We do this for the case of  $Y^{p,k}(\C P^2)$. We have
\begin{equation}
\begin{array}{lcll}
Y^{p,3p}(\C P^2)  ~: & & \vol (\Xi_2) \,=\, \vol(\Theta_i) \,= \,\frac{\pi^3}{3p} \nn\\[2.5mm]
Y^{2r,3r}(\C P^2) ~: & & \vol (\Xi_1) \,= \, \vol (\Xi_2) \, = \, \frac{9\pi^3}{64r}~, \qquad \vol(\Theta_i) \,= \,\frac{3\pi^3}{16r}~.
\end{array}
\end{equation}
Notice that in the case $Y^{p,3p}(\C P^2) = S^7/\Z_{3p}$ the volume of $\Xi_1$ is formally zero. The fact that one submanifold
disappears in this limit  may be also understood from the fact that the number of external points in the toric diagram jumps from five to four, as
discussed  around equation (\ref{5legs}).

Notice that the volumes given above satisfy the relation
\bea
\sum_{i=1}^2 \vol (\Xi_i) + \sum_{i=1}^g \vol (\Theta_i)  \, = \,  \frac{12}{\pi}\,\vol (Y_{7}) ~.
\eea
In fact, this follows from specialising a general formula (\emph{cf}. equation (2.88))
given in the first reference in \cite{zmin}.  Using (\ref{condim}), with $N=1$, this may be rewritten as
\bea
\sum_{a=1}^{d} \Delta_a \, =\, 2 ~.
\eea
In the context of AdS$_5$/CFT$_4$ this formula is interpreted as the constraint that 
the R-charges of the fields entering in a superpotential term sum to two \cite{Franco:2005sm}, 
and it is natural to give the same interpretation in the context of AdS$_4$/CFT$_3$ .

%%%%%%%%%%%%%%%%%%%%%%%%%%%%%%%%%%%%%%%%%%%%%%%%%%%%%%%%%%%%%%%%%%%%%%%%%%%%%%%%%%%%%%%%%%%%%%

\section{Supergravity solutions}
\label{sugrasection}

We now turn to a discussion of the AdS$_4\times Y_7$ M-theory backgrounds and their reduction to type IIA string theory.
We then describe in more detail the orbifold $S^7/\Z_{3p}$ and its cone $\C^4/\Z_{3p}$.

%%%%%%%%%%%%%%%%%%%%%%%%%%%%%%%%%%%%%%%%%%%%%%%%%%%%%%%%%%%%%%%%%%%%%%%%%%%%%%%%%%%%%%%%%

\subsection{M-theory and type IIA backgrounds}

We use the notation of \cite{Gauntlett:2005jb,ABJM}. The 
M-theory backgrounds of interest take the form
\bea
\diff s^2 & = & R^2 \left(\frac{1}{4} \diff s^2 (\mathrm{AdS}_4) + \diff s^2(Y_7)\right)~,\nn\\
G_4 & = & \frac{3}{8} R^3 \diff \vol (\mathrm{AdS}_4)   ~,
\eea
where the Einstein metrics on AdS$_4$ and $Y_7$ obey
\bea
\mathrm{Ric}_{\mathrm{AdS}_4} \, =\,  3 \, g_{\mathrm{AdS}_4}\qquad \quad \mathrm{Ric}_{Y_7} \, =\, 6 \, g_{Y_7}~,
\eea
respectively. The radius $R$ is determined by the quantisation of the $G_4$ flux
\bea
N \, = \, \frac{1}{(2\pi l_p)^6} \int_{Y_7} * G_4~,
\eea
where $l_p$ is the eleven-dimensional Planck length, given by
\bea
R^6 \, =\, \frac{(2\pi l_p)^6 N}{6\vol (Y_7)} ~.
\label{bigradius}
\eea

Recall that Sasaki-Einstein metrics may be canonically written as 
\bea
\diff s^2(Y_7) \, =\, \diff s^2 (B_6) + (\diff \varphi + \sigma )^2 ~,
\label{reebfib}
\eea
where $\diff s^2 (B_6)$ is in general only a local K\"ahler-Einstein metric (with $\mathrm{Ric}_{B_6} \, =\, 8 \, g_{B_6}$) and
$\diff \sigma /2 = \omega_{B_6}$ is the corresponding K\"ahler two-form. When the Sasaki-Einstein manifold $Y_7$ is 
of (quasi-) regular type, meaning that $B_6$ is a manifold (orbifold),  one may then quotient by the 
$U(1)$ action generated by the Reeb vector field $\de_\varphi$. Thus, in these cases one can reduce to type IIA supergravity along
this particular direction. This is the reduction discussed in \cite{ABJM} for the case of $Y_7=S^7$, or more generally 
$Y_7=S^7/\Z_k$. The $\Z_k$ action discussed by ABJM divides by a factor of $k$ the periodicity of $\varphi$.
 The radius of the M-theory circle is in this case $R_\varphi = R/k\sim (N/k^5)^{1/6}$, and 
thus the M-theory description is valid for $N>> k^5$ \cite{ABJM}.  On the other hand, when $N<< k^5$ 
the circle becomes small and one should pass to a type IIA description.
The resulting type IIA supergravity solution preserves ${\cal N}=6$ supersymmetry at the supergravity level 
\cite{Sorokin:1985ap}, and gives the background
\bea
\diff s^2_{st} \, = \, \frac{R^3}{k} \left(\frac{1}{4} \diff s^2 (\mathrm{AdS}_4) + \diff s^2(\C P^3)\right)~,\qquad \quad\\[2mm]
\ex^{2\Phi} \, = \, \frac{R^3}{k^3}~, \qquad F_4 \, = \,  \frac{3}{8} R^3 \diff \vol (\mathrm{AdS}_4)~, \qquad 
F_2 \, = \, 2 k \, \omega_{\C P^3}~,
\eea
where the metric is in the string frame. There are then $N$ units of $F_4$ flux through AdS$_4$, and $k$ units of $F_2$
flux through the linearly embedded $\C P^1 \subset \C P^3$. Note here that, due to the normalisation of 
the K\"ahler-Einstein metric on $\C P^3$, the Ricci form of the latter is given by $\rho = 8\omega_{\C P^3}$, 
and thus
\bea
\int_{\C P^1} \frac{\omega_{\C P^3}}{2\pi} \, =\, \frac{1}{8}\int_{\C P^1} c_1(\C P^3)\, =\, \frac{1}{2}~.
\eea
Here we have used the fact that the first Chern class of the tangent bundle of $\C P^3$ is 
equal to $4$ times the hyperplane class.
The radius of curvature of this background is $R^2_{st}= R^3/k \sim (N/k)^{1/2}$, 
and thus the type IIA supergravity approximation  is valid for $N>> k$ \cite{ABJM}.

One might consider performing a similar reduction of one of the homogeneous Sasaki-Einstein manifolds 
to a solution of type IIA supergravity. However, because one starts with ${\cal N}=2 $ supersymmetry only, now all supersymmetries 
are broken in the reduction  \cite{Sorokin:1985ap}.  Moreover, for generic $Y^{p,k}_7$ manifolds, there is no way to make sense of the quotient 
space, even locally, as a manifold.

However, from the construction of \cite{paper3} reviewed in section \ref{revpaper3},
we see that one may consider a different reduction along the $\alpha$-circle, obtaining perfectly smooth ${\cal N}=2$ 
supersymmetric\footnote{This follows since both Killing spinors of the Sasaki-Einstein seven-manifolds 
are invariant under this $U(1)_\alpha$ action. See \emph{e.g.} \cite{Gauntlett:2005jb} for an explicit calculation.}
type IIA backgrounds. These are \emph{warped} products AdS$_4\times M_6$, with RR fields and a non-trivial dilaton. 
The manifolds $M_6$ are  $S^2$ bundles over the K\"ahler-Einstein manifold $B_4$.
In fact this bundle is obtained from the canonical bundle\footnote{However, one should note that the
natural complex structure here is \emph{different} from the one
associated to the Calabi-Yau cone.} $\mathcal{L}$ over $B_4$
by replacing the $\C$ fibre by $\C P^1$. Note that in
\cite{paper3} it was shown that $M_6$ are always spin manifolds. 
 The topology of $M_6$ was discussed earlier. 
To perform the reduction we write $\diff s^2 (Y_7) = \diff s^2 (M_6) + w(\rho)\ell^2 (\diff \gamma + \ell^{-1}B)^2$, obtaining
\bea
\diff s^2_{st} \, = \, \sqrt{w(\rho)}\, \ell R^3 \left(\frac{1}{4} \diff s^2 (\mathrm{AdS}_4) + \diff s^2(M_6)\right)\qquad \quad\\[2mm]
\ex^{2\Phi} \, = \, \ell^3 R^3 (w(\rho))^{3/2} \qquad F_4 \, = \,  \frac{3}{8} R^3 \diff \vol (\mathrm{AdS}_4) \qquad F_2 \, = \, \ell^{-1}\diff B~,
\eea
where $w(\rho)= (1-8\rho^2/3 +\kappa / (48\rho^4) )/16$ is a bounded function on $M_6$. 
From section (\ref{revpaper3})  we find that the RR two-form flux has quantised periods, 
namely 
\bea
\int_{\Sigma } \frac{F_2}{2\pi} \, =\, p ~, \qquad  \qquad \quad \int_{\sigma^N \Sigma_i} \frac{F_2}{2\pi} \, =\, k~.
\label{F2fluxes}
\eea
Here $\Sigma \cong S^2$, and  $\sigma^N \Sigma_i$ is either a copy of $\C P^1 \subset \C P^2$, or one 
of the 
two copies of $\C P^1 \subset \C P^1 \times \C P^1$, in the two examples, respectively. 
Notice that $\kappa\sim 1$, so $w(\rho)$ is of order 1 in $p$ and $k$. The radius of the M-theory circle is 
\bea
R_\gamma \,=\,\ell  R \, \sim \, \frac{\ell N^{1/6}}{\vol (Y_7)^{1/6}} ~,
\eea
and we should pass to a type IIA description when this is small. 
The radius of curvature in the type IIA solution is 
\bea
R^2_{st} \, =\, \ell R^3 \, \sim \, \frac{\ell N^{1/2}}{\vol (Y_7)^{1/2}}~.
\eea
Recall that $\vol (Y_7)$ and $\ell$ are determined in terms of $p$ and $k$ through (\ref{volumefor}),  (\ref{alpharadius}),
and the range of $k$ is  constrained by the value of $p$. 
Thus, we can consider the limit  $p>>1$, $k>>1$, at fixed $p/k$. 
Since $x_i\sim 1$, we have both $\ell \sim 1/p$, $\vol (Y_7)\sim 1/p$, thus we obtain a behaviour 
qualitatively similar to the orbifold case, reviewed above. 
In particular, the M-theory description is valid when $N >> p^5$, while type IIA
supergravity is a good approximation in the regime $p^5  >>  N >> p $.

\subsection{The orbifolds $S^7/\Z_{3p}$ and $\C^4/\Z_{3p}$}

For the five-dimensional $Y^{p,q}$ manifolds, understanding the limiting case 
$Y^{p,p}=S^5/\Z_{2p}$ was a key step for constructing the  complete family of quiver gauge 
theories \cite{quivers}.
We hence now discuss in more detail the analogous case of the $Y^{p,3p}=\C^4/\Z_{3p}$ orbifold. 

In terms of standard complex coordinates on $\C^4$, the orbifold action (\ref{orbi3p}) is
\bea
(z_1, z_2, z_3, z_4) \quad \to\quad (\ex^\frac{2\pi i}{3p} z_1 , \ex^\frac{2\pi i}{3p} z_2,
\ex^\frac{2\pi i}{3p} z_3, \ex^{-\frac{2\pi i}{p}} z_4) ~.
\eea
The orbifold therefore preserves ${\cal N}=2$ supersymmetry \cite{Benna:2008zy,Terashima:2008ba}. 
We begin by noting that after the following non-holomorphic change of coordinates
\bea
w_1=z_1, \quad w_2=z_2, \quad w_3=z_3, \quad w_4=\bar{z_4}~,
\eea
the above orbifold acts as the $3p$-th roots of unity $\Z_{3p}\subset U(1)$ acting on
$\widetilde{\C^4}$ with weights $(1,1,1,3)$. The quotient by the latter realises $S^7$ as a
$U(1)$ orbi-bundle over the weighted
projective space $W\C P^3_{[1,1,1,3]}$. For $p=1$, one divides by $\Z_3$ along the fibre,
resulting in the
solution AdS$_4\times W\C P^3_{[1,1,1,3]}$, with three units of of RR $F_2$ flux  through
the $\C P^1$ and  one unit of flux at a $\Z_3$ orbifold singularity -- see (\ref{F2fluxes}).
 For general $p$ these are replaced by $3p$ units and $p$ units, respectively. 

We may also understand this orbifold via the 
canonical Hopf fibration (\ref{reebfib}) of $S^7$ over $\C P^3$. 
Note that the orbifold acts as a subgroup of $SU(4)$ acting on $\C^4$,
which descends to an action on $\C P^3$ itself. For simplicity, we discuss the case $p=1$ -- 
the general $p>1$ case is a further $\Z_p$ quotient of this geometry. 
The action is
\bea
(\omega_3,\omega_3,\omega_3,1)~.
\eea
On $\C^4$ this fixes the complex line $(0,0,0,z_4)$. The action on the copy of 
$\C^3$ given by $(z_1,z_2,z_3,0)$ is the usual diagonal Lens space action, 
with the $\Z_3\subset U(1)$  acting along the Hopf fibre of $S^5\rightarrow \C P^2$. 
The action is thus 
free away from the origin. 
We now descend to $\C P^3$. 
We obtain in this way a $U(1)$
bundle over $\C P^3/\Z_{3}$. 
The orbifold action has fixed points at a point 
and the linearly embedded $\C P^2$. Indeed, where $z_4\neq 0$ we may introduce 
homogeneous coordinates
\bea
x_1=\frac{z_1}{z_4}, \quad x_2 = \frac{z_2}{z_4},\quad x_3=\frac{z_3}{z_4}~.
\eea
The $\Z_3$ action is simply the diagonal action, which thus has an isolated 
$\Z_3$ fixed point $\{x_1=x_2=x_3=0\}$. Similarly, the $\C P^2$ at $z_4=0$ 
is also fixed by the orbifold action. In fact the orbifold action acts 
on the Hopf fibre over this $\C P^2$, as mentioned above. Thus the $U(1)$ bundle restricted to 
$\C P^2$ is $O(-3)$. 

The resulting orbifold of $\C P^3$ may be viewed as follows. 
We begin by viewing $\C P^3$ as $O(1)_{\C P^2}$ glued to 
an open ball in $\C^3$ -- both have boundary $S^5$. We may 
also think of this as collapsing the boundary of 
$O(1)_{\C P^2}$ to a point $p_\infty$. This is the point $\{x_1=x_2=x_3=0\}$ above. 
The $\Z_3$ action 
is along the fibre of $O(1)_{\C P^2}$, which is also the Hopf fibre of 
the $S^5$. Thus we see explicitly that the $\C P^2$ zero section and the 
point $p_\infty$ are fixed. We may construct the same space by instead
starting with $O(3)_{\C P^2}$. The boundary is $S^5/\Z_3$, which collapsing 
to a point in the same way means that $p_\infty$ is now an isolated $\Z_3$ singularity. Note that 
originally the $\C P^2$ zero section was a fixed locus of the $\Z_3$ action. However, 
$\C/\Z_3\cong \C$, and thus the two spaces we have described are diffeomorphic, 
although not equivalent as orbifolds. 

The discussion in the above paragraph is precisely analogous
to the discussion of the orbifold $S^5/\Z_2$ in \cite{KW}. Following 
the latter reference, we may thus ask what happens when 
we blow up the isolated $\Z_3$ singularity at the point $p_\infty$. 
This results in the space
\bea\label{M6space}
\C P^1 \times_{U(1)} O(3)_{\C P^2}~.
\eea
This is a $\C P^1$ bundle over $\C P^2$, and in fact is precisely the base space $M_6$ 
in the construction of section \ref{firstsection}. However, unlike 
\cite{KW}, we cannot interpret this as the base of the homogeneous 
space $M^{3,2}$, since (\ref{M6space}) is not diffeomorphic to 
$\C P^1\times \C P^2$. This suggests that we cannot view 
the $M^{3,2}$ theory as the IR fixed point of a deformation 
of the orbifold $S^7/\Z_3$, in the same way that $T^{1,1}$ arises 
as a relevant deformation of $S^5/\Z_2$ \cite{KW}. 

It is also clear in this description that the four supersymmetric
 5-submanifolds are in this case copies of $S^5/\Z_{3p}$. Note 
that one of these is a smooth Lens space, with action generated by $(\omega_{3p},\omega_{3p},\omega_{3p})$, 
whereas the other three are isomorphic to each other, 
being \emph{singular} quotients $(\omega_{3p},\omega_{3p},\omega^{-3}_{3p})$. 
In fact these latter quotients are similar 
to the $S^5/\Z_2$ quotient, mentioned above.

Note that when $p$ is even the orbifold action contains elements that 
act diagonally along the Hopf $U(1)$. To see this, note that 
the condition for an element to act along the Hopf diagonal is
\bea
\frac{l}{3p} \cong -\frac{l}{p}\quad \mathrm{mod} \ 1
\eea
which implies
\bea
4l = 3pn
\eea
where, without loss of generality, we take $0<l<3p$ so that $n\in \{1,2,3\}$. 
Clearly for $p$ odd this has no solution. However, for $p=2r$ even 
we may in general take $l=3r$, $n=2$, which leads to the 
diagonal $\Z_2$ action on $\C^4$
\bea
 (z_1,z_2,z_3,z_4)\, \to \, - (z_1,z_2,z_3,z_4)~.
\label{abjmlev2}
\eea
This is precisely the $k=2$ orbifold action considered by ABJM \cite{ABJM}. 
On the other hand, if $p$ is divisible by $4$, so $p=4m$, we may 
take $l=3m$, $n=1$, leading to the diagonal $\Z_4$ action generated by
\bea
(z_1,z_2,z_3,z_4)\, \to \, \omega_4 \cdot (z_1,z_2,z_3,z_4)~.
\eea
This is the $k=4$ orbifold action considered by ABJM. In these
latter two cases we may view the orbifold instead as 
$(\C^4/\Z_2)/\Z_{3r}$ and $(\C^4/\Z_4)/\Z_{3m}$, respectively, 
where the first quotient is the ABJM quotient.

Notice that in the discussion above one has to be careful 
about which complex structure one is using on $\C^4$. Recall that  the $\Z_k$ action considered by ABJM is actually 
a discrete subgroup of the baryonic $U(1)_B$,  
acting  as follows on the bifundamental fields 
\bea
 A_i \to \ex^{i \alpha} A_i~, \quad \qquad B_i \to \ex^{-i\alpha} B_i~.
\eea
Setting $\alpha = 2\pi/k$, we see that  $\Z_k \subset U(1)_B$. Thus, on the natural 
GLSM coordinates\footnote{The GLSM description gives the conifold as a $\C^4//U(1)_B$ quotient.} 
 $z_i$ on $\C^4$, the ABJM $\Z_k$ quotient acts as
\bea
(z_1,z_2,z_3,z_4) \to ( \ex^{i 2\pi/k}  z_1, \ex^{i 2\pi/k} z_2,  \ex^{-i 2\pi/k} z_3, \ex^{-i 2\pi/k} z_4)~.
\eea
The coordinates on $\C^4$ used in \cite{ABJM} are  related to the above coordinates by  a non-holomorphic change of variable:  $z_1'= z_1$, $z_2'=z_2$, $z_3'=\bar z_3$, $z_4'=\bar z_4$. Notice that for $k=2$ (and only for this value) 
the action on $z_i$ and $z_i'$ is obviously the same.
To construct $\mathcal{N}=2$ orbifold quivers of the ABJM theory, it seems more appropriate to use the 
orbifold action on the $z_i$ coordinates above. However, it is not clear that the standard rules 
(\cite{DM}) for constructing 
four-dimensional orbifold quivers will apply.  

%%%%%%%%%%%%%%%%%%%%%%%%%%%%%%%%%%%%%%%%%%%%%%%%%%%%%%%%%%%%%%%%

\section{Discussion}
\label{disc}

In this paper we have studied in detail two of the families of Sasaki-Einstein 
seven-manifolds constructed in \cite{paper3}. These are the simplest examples, with 
the largest isometry groups. In particular, we have given gauged linear 
sigma model descriptions of these manifolds, discussed their topology, and also 
described relevant supersymmetric submanifolds and their volumes. 
As is the case for the five-dimensional $Y^{p,q}$ manifolds \cite{toric}, 
we have shown that these families interpolate between certain orbifolds 
of homogeneous Sasaki-Einstein manifolds. In particular, 
the family $Y^{p,k}(\C P^2)$ has a limit $Y^{p,3p}(\C P^2)=\C^4/\Z_{3p}$, 
and we discussed this orbifold in some detail. 
The geometric results of this paper should be a useful first step 
in constructing candidate AdS$_4$/CFT$_3$ dual superconformal field theories. 
We conclude by discussing some of the issues involved in pursuing this 
programme. 

As a general comment, note that a key ingredient in AdS$_5$/CFT$_4$ duality 
involving Sasaki-Einstein five-manifolds is $a$-maximisation \cite{IW}. Among the consequences of 
$a$-maximisation
is the fact that the central charges, as well as the R-charges of a given SCFT,
 are necessarily algebraic numbers, \emph{i.e.}
roots of polynomials with integer coefficients. It was proven in 
\cite{zmin} that the volumes, and volumes of supersymmetric submanifolds, 
of Sasaki-Einstein manifolds are always algebraic numbers, in 
\emph{any} dimension. For the examples discussed in this paper 
we obtain cubic irrational numbers. This strongly suggests that
there should be some type of analogue of $a$-maximisation for
three-dimensional conformal field theories with
${\cal N}=2$ supersymmetry. Note that the field theoretic $\tau$-minimisation of 
\cite{Ken} applies to such theories, although it is currently not known 
how to use this to obtain exact field theory results.

The Calabi-Yau cones $C(Y^{p,k}_7)$ we have discussed admit explicit 
Calabi-Yau resolutions, or partial resolutions where there 
are residual orbifold singularities \cite{resolutions}. 
This fact might be useful for obtaining further insight into these 
theories \cite{resolving1,resolving2}.
Note that such resolutions would also allow the BPS ``mesonic'' spectrum to be 
read off  \cite{bpsmesons} from the index-character of \cite{zmin}. 
Indeed, such generating functions have already been computed 
for the handful of currently-known orbifold duals in \cite{Ami}.

Since the geometries are toric, there will also be 
a dual brane web description. In this case the
Calabi-Yau cones may be described as Special Lagrangian
$\T^3\times\R$ fibrations over $\R^4$, with certain types
of degeneration of the fibres encoded combinatorially in
terms of toric data. Reduction and two T-dualities leads to
a dual description in terms of prq-4-branes in type IIA \cite{LV}.
The configuration of these 4-branes may be
read off from the toric data we presented earlier.
This leads to a three-dimensional ``web diagram'', describing the
locus of the prq-4-branes. The problem of finding the dual
gauge theory then becomes translated into a problem of understanding
the effective theory of such webs of 4-branes. 
Again, the toric nature of these manifolds also implies that 
one can write down associated M-theory crystals \cite{crystal}. 
These are analogues of dimer configurations, although 
it is not clear to us how these are related to the recent 
Chern-Simons gauge theory construction of \cite{ABJM}, and 
various follow-up papers.

A possible  avenue of research is to try to 
construct a Chern-Simons-matter theory that is dual 
to the orbifold $\C^4/\Z_{3p}$. Similar orbifold 
theories have recently been constructed and discussed in \cite{Benna:2008zy, 
Imamura:2008nn, Terashima:2008ba, Aharony:2008gk}. 
This should be, in some sense, a limiting theory of 
the theories dual to $Y^{p,k}(\C P^2)$. The aforementioned
orbifold constructions simply apply the standard 
methods  to construct the orbifold theories. 
However, the reasoning for this is currently obscure. 
In particular, the ABJM orbifold $S^7/\Z_k$ is 
not simply a standard orbifold projection 
of the theory for $k=1$ -- instead one changes the 
Chern-Simons level from $k=1$ to $k$. A systematic
understanding of how to construct orbifold 
theories is currently lacking. However, note 
that a necessary condition for a candidate theory 
to be dual to a particular AdS$_4 \times Y_7$ background 
is that its vacuum moduli space contains 
the $N$th symmetric product of $C(Y_7)$ 
as a subvariety. This is because the latter
is the moduli space of $N$ M2-branes that 
are transverse to the Calabi-Yau singularity
$C(Y_7)$. This problem, for general classes
of $d=3$, $\mathcal{N}=2$ Chern-Simons quiver gauge theories, will be 
addressed in \cite{VMSpaper}.

\subsection*{Acknowledgments}
\noindent
D. M.  acknowledges support from NSF grant PHY-0503584. J. F. S. 
is funded by a Royal Society University Research Fellowship.

\end{document}